\documentclass{eptcs}
\usepackage{alltt}

\providecommand{\event}{LAFM 2013} 

\title{Analyzing Behavioural Scenarios over Tabular Specifications Using Model Checking}
\def\titlerunning{Analyzing Behavioural Scenarios over Tabular Specifications Using Model Checking}

\author{Gast\'on Scilingo \qquad Mar\'{\i}a Marta Novaira \qquad Renzo Degiovanni\textit{*}
\institute{Departamento de Computaci\'on, Universidad Nacional de R\'{\i}o Cuarto, Argentina}
\institute{\textit{*}Consejo Nacional de Investigaciones Cient\'{\i}ficas y T\'ecnicas (CONICET), Argentina}
\email{\{gaston, mnovaira, rdegiovanni\}@dc.exa.unrc.edu.ar}
}
\def\authorrunning{G.~Scilingo, M.~M.~Novaira and R.~Degiovanni}
\begin{document}
\maketitle

\begin{abstract}
Tabular notations, in particular SCR specifications, have proved to be a useful means for formally describing complex requirements.
The SCR method offers a powerful family of analysis tools, known as the SCR Toolset, but its availability is restricted by the Naval Research Laboratory of the USA. This toolset applies different kinds of analysis considering the \emph{whole} set of behaviours associated with a requirements specification.

\noindent
In this paper we present a tool for describing and analyzing SCR requirements descriptions, that complements the SCR Toolset in two aspects. First, its use is not limited by any institution, and resorts to a standard model checking tool for analysis; and second, it allows one to concentrate the analysis to particular sets of behaviours (subsets of the whole specifications), that correspond to particular scenarios explicitly mentioned in the specification.

\noindent
We take an operational notation that allows the engineer to describe behavioural ``scenarios'' by means of programs, and provide a translation into Promela to perform the analysis via Spin, an efficient \emph{off-the-shelf} model checker freely available.
In addition, we apply the SCR method to a \emph{Pacemaker} system and we use its tabular specification as a running example of this article.

\end{abstract}

\section{Introduction}

It is generally accepted that the quality of requirements specifications has a great impact in the whole development process~\cite{Ghezzi-FSE+2002, Jalote2005, Sommerville2006}. Requirements specifications are mostly expressed informally. Various approaches deal with this informal representation, helping in eliciting, expressing and organizing requirements, via languages such as DFDs and use cases~\cite{alexander2004scenarios}. However, \emph{formal} requirements specifications are better suited for analysis and its automation. Tabular notations, originally used to document requirements by D.~Parnas and others \cite{Heninger1978}, have proved to be a useful means for concisely and formally describing expressions characterizing complex requirements. Tables are used for describing \emph{relations} between the monitored and controlled environment variables, and assumptions over the environment. Tables impose a \emph{structure} on specifications, which helps in organizing large and complex formulas into well structured smaller formulas that are easier to follow, and imposes simple constraints for guaranteeing completeness (no missing cases) and consistency (no contradicting requirements).
Tabular requirements notations evolved to become the Software Cost Reduction (SCR) method \cite{HeitmeyerLK95}, which has been widely used to describe software requirements \cite{Butler96anintroduction, Heitmeyer1997LFM, Heninger1978}. The SCR Toolset assists this method, by providing several tools to help the engineer in the task of assessing the quality of an SCR requirements specification~\cite{csse-HeitmeyerABJ05}. It provides different kinds of useful analyses: type-checking, consistency checking, simulation of user provided scenarios, user assisted property checking based on theorem proving, and the automated analysis of certain properties via model checking. However, this toolset has two major drawbacks. First, it belongs to the Naval Reseach Laboratory of the United States, and its availability is controlled by this institution (the toolset is not freely available); and second, most analyses supported by the SCR Toolset apply to the \emph{whole} set of behaviours of the specification (with the sole exception of the scenario simulation tool). This, of course, seriously limits the use of SCR as a formal method for specifying requirements, despite its proved success in the specification of various critical systems.

In this paper we present a tool for describing and analyzing behavioural scenarios over SCR specifications. We take an operational notation that allows the SCR engineer to describe behavioural ``scenarios'' by means of programs, and provide a translation of these programs into Promela, to perform analysis via Spin, an efficient \emph{off-the-shelf}, freely available, model checker. Essentially, these programs over tables provide the means for \emph{guiding} the execution of the specified system, and therefore to express properties that may be of interest only for these ``guided'' scenarios. The relevance of this mechanism is justified by the fact that many requirements are typically expressed in terms of particular scenarios. 
The relevance of this kind of requirements is also evidenced by the various ``scenario-based'' notations that are typically used for requirements specification, such as use cases, activity diagrams, and message-sequence charts. 


We introduce our tool by specifying and analyzing particular critical scenarios for the SCR specification of a \emph{Pacemaker} system. A \emph{Pacemaker} is a highly critical system and possibly more complex that the frequent examples found in the SCR literature. The selection of this case study was motivated by the Formal Methods Challenge of the Software Quality Research Laboratory of McMaster University \cite{Pacemaker-Spec}. We also evaluate our tool by comparing it against a previous analysis approach that also supports scenarios over tabular specifications.

\section{Motivating Example: A description of a Pacemaker}
\label{pacemaker}

Let us briefly describe the \textit{Pacemaker} system, the example that we use as a vehicle to introduce the notation and the tool. The full informal description can be found in~\cite{Pacemaker-Spec}.
A Pacemaker monitors and regulates a patient's heart rate. The device is conformed by three parts: the \textit{Pulse Generator} (PG) produces programmable pulses (atrial and ventricular) which provide electrical stimulation to the heart for pacing; \textit{Device Controller-Monitor} (DCM) allows the physician to program and consult the PG; and the \textit{leads} implanted in the patient allows the device to sense the heart's activity and delivers pacing therapy to the patient's heart.

SCR, as other formal requirements specification methods, requires one to identify monitored and controlled variables (those the system depends on, and controls as part of its behaviour, respectively). The Pacemaker depends on the following monitored variables: the battery voltage level (\texttt{mBATTERYvoltage}), the commands that PG receives from the DCM (\texttt{mCommand}), and the program configured by the physician for the bradycardia therapy (\texttt{mMODEbrad}). In addition, the system controls variables that indicate the chamber to be paced and sensed. For instance, if the controlled variable \texttt{cCHAMBERSpaced} has the value \texttt{A}, then PG should pace the atrial chamber.

The system internal state is captured via a state variable called mode class, and by auxiliary (intermediate) variables called terms. From the informal description, we distinguished five different bradycardia states that conform the unique system mode-class (\texttt{mcPulseCondition}). In \texttt{Normal} mode the device provides the patient with the therapy programmed by the physician. The \texttt{Temporal} mode is used to temporarily test various system parameters.
The mode \texttt{PaceNow} is an emergency bradycardia pacing commanded by the physician. A sample term is \texttt{tMagnetON}, which captures the fact that the magnet is near enough to the PG. The magnet affects the state of the system: the magnet mode is used to test the battery status of the device. When this state is released, the PG should return to the previous bradycardia state. In order to ``remember'' the mode to go back to, the magnet mode is refined into \texttt{MAGnormal}, \texttt{MAGpacenow}, \texttt{MAGpor} and \texttt{MAGtemporal}. 
Finally, the Power on Reset mode (\texttt{POR}) should be entered when the battery voltage drops lower than a threshold (\texttt{BatteryLevel}), when the PG operation becomes unpredictable. All functions are disabled until the battery voltage exceeds the threshold, optimizing the battery usage and ensuring a minimal pulsing to the patient. Due to space restrictions, we only show a small fragment of the mode transition table, a table that specifies how the system changes its mode as a consequence of events. The notation \texttt{@T}, \texttt{@F} and \texttt{@C} is used to indicate that a formula (which can be a state formula or an event) becomes true, false or changes its value, respectively. For instance, @T(mBATTERYvoltage $<$  BatteryLevel) expresses that the battery voltage becomes lower than the battery level. 

\begin{figure}[t]
{\scriptsize
\texttt{
\begin{center}
\begin{tabular}{|l|l|l|}
\hline
OLD MODE & EVENT & NEW MODE\\
\hline
Normal & @T(mBATTERYvoltage $<$  BatteryLevel) & POR\\
\hline
Normal & @T( tMagnetON ) when mMagnet=ON & MAGnormal\\
\hline
POR & @T( mCommand=NORMAL ) & Normal\\
\hline
MAGnormal & @F( tMagnetON ) & Normal\\
\hline
MAGpor & @F( tMagnetON ) & POR\\
\hline
..... & ..... & .....
\end{tabular}
\end{center}
}
\caption{Fragment of the modes transition table for \texttt{mcPulseCondition}.}%
\label{tab:mode-table}%
}
\end{figure}

The complexity and criticality of a Pacemaker makes it essential to verify certain properties of the system. Particular properties of interest are: \emph{``When the magnet state is released, the PG should return to the previous bradycardia state''}, and \textit{``When the battery voltage drops lower than a threshold, the Power-on-reset (POR) state should be entered due to the system operation being unpredictable. All functions remain disabled until the battery voltage exceeds that threshold''}. Notice that these properties correspond to particular scenarios of the system.

\section{Specifying Behavioural Scenarios}
\label{notation}

In many situations the engineer is interested in analyzing particular sets of executions, that correspond to scenarios detected from the (informal) requirements description. Let us describe an operational notation for describing \emph{scenarios} over tabular specifications, a variant of that originally introduced in \cite{DBLP:conf/fase/AguirreFMMW09}. Scenarios will not be individual executions (as they are in the case of the simulation tool of SCR Toolset). They will be families of executions, described in terms of programs referring to the tabular descriptions. 
The syntax for specifying the scenarios is the following:
\begin{quote}
{\small
\begin{alltt}
\textit{SCENARIO} ::= program : { \textit{PROGRAM} } check : \textit{PROP}
\textit{PROGRAM}  ::= \textit{SENTENCE} | \textit{SENTENCE} ; \textit{PROGRAM} | \textit{PROGRAM}*
\textit{SENTENCE} ::= [ \textit{FORM} ] | stateChange | stateChange[ \textit{FORM} ]
\end{alltt}
}
\end{quote}

\noindent
An annotated \emph{scenario} is a program and a property to be checked (a state formula that should hold at the end of the execution). A \emph{program} is composed of \emph{sentences} combined via sequential compositions (\texttt{;}), iterations (\texttt{*}), and other constructs.
A \emph{test} sentence is an expression \texttt{[ f ]}, where \texttt{f} is a state formula, that represents a transition that does not modify the state but can only be executed when \texttt{f} is true. The \texttt{stateChange} sentence represents an arbitrary (non deterministic) atomic \emph{change} in the state. The restricted change sentence \texttt{stateChange[f]} allows us to ``guard'' a state change, by either an event or a condition. 
Consider for instance the following sentences: 
{\small
\begin{alltt}
\textit{(i)} stateChange[NOT @T(tMagnetOn)] \textit{(ii)} stateChange[@T(mBATTERYvoltage < BatteryLevel)] 
\textit{(iii)} [mcPulseCondition = MAGnormal]
\end{alltt}
}

\noindent
The first one models any change in which the magnet is not close; the second expresses that the change is a drop in the battery voltage; and the last one is a test sentence that is executed only if the current mode is \texttt{MAGnormal}. 
Let us now specify, via programs, one of the above scenarios identified for the Pacemaker system.

{\small
\begin{verbatim}
program : { 
  stateChange*; [ mcPulseCondition = MAGnormal ];  stateChange[@F(tMagnetON)] 
  }
check : { mcPulseCondition = Normal }
\end{verbatim}
}

\noindent
This program makes zero or more unrestricted changes (executing \texttt{stateChange*}), until the \texttt{MAGnormal} mode is reached. Then the magnet is removed (\texttt{stateChange[@F(tMagnetON)]}). We would want to check if the PG finalizes in \texttt{Normal} mode, the previous mode before entering magnet mode \texttt{MAGnormal}. 

\section{Analyzing Behavioural Scenarios}
\label{analyzing-scenarios}

In order to be able to verify properties of behavioural scenarios, the tool needs to capture SCR tables in Promela. This is done as put forward by Bharadwaj and Heitmeyer~\cite{ase-BharadwajH99}. Our tool then proceeds to automatically encode programs into the obtained Promela model, so that the resulting model can be analyzed via model checking.
Due to space restrictions, we only highlight the important points that the translation performed by our tool takes into consideration. Let $a_1; a_2; \ldots; a_n$ be a specified scenario. First, the tool enumerates each sentence in the scenario with a ``program counter'' ($pc$) that represents the progress in the execution. For each sentence $a_i$, the translation proceeds as follows:
\begin{itemize}
\item if $a_i$ is \texttt{stateChange}, then no encoding is needed because the tables characterization already take care of arbitrary changes in the state.
\item if $a_i$ is the restricted change \texttt{stateChange[ f ]}, we have to preserve the transitions that satisfy \texttt{f} and remove those that do not.
\item if $a_i$ is test \texttt{[ f ]}, the treatment is similar to restricted changes but it does not produce a new state. It filters traces in which the current state does not satisfy \texttt{f}.
\item if $a_i$ is an iteration, a special treatment is considered. We use the Promela non-deterministic assignment to model that $a_i$ is executed zero, one or more times.
\end{itemize}

\noindent
The notation introduced allows us to specify the property to be checked at the end of the scenario. As each sentence in the scenario has a program counter assigned by the tool, the property has to hold at the end of the execution (i.e, when $pc = n+1$). Then, the tool for verifying the property $PROP$ generates the following Promela assertion:
\texttt{assert(pc==n+1 $\rightarrow$ \textit{PROP})}.

\section{Experimental Results}
\label{experimental-results}

We show two kinds of experiments in this section. First, we describe and analyze particular behavioural scenarios for the Pacemaker SCR specification. We assess the ability of the tool in finding problems in the specification, and how this helps us improving the description. Second, we evaluate our tool by comparing it against a previous approach introduced in~\cite{DBLP:conf/fase/AguirreFMMW09} that also supports scenarios over tabular specifications, but the analysis is performed by DynAlloy Analyzer~\cite{icse-FriasGPA05}.
%
In section~\ref{pacemaker} we mentioned two scenarios that we had identified from the informal description of a Pacemaker system:

\begin{description}
\item ($S1$) \emph{``When the battery voltage drops lower than a threshold, the Power-on-reset (POR) state should be entered due to the system operation being unpredictable. All functions remain disabled until the battery voltage exceeds that threshold''}
\item ($S2$) \emph{``When the magnet state is released, the PG should return to the previous bradycardia state.''}. 
\end{description}

\noindent
Scenario \textit{S1} describes one of the most critical situations in a Pacemaker system. While \textit{S2} models a situation that happens when the physician configures the PG using the DCM. Surprisingly and contrary to our expectations, our tool was able to find a counterexample for both scenarios (i.e., the model checker found violations to both properties).  The counterexample for \textit{S1} has to do with the system changing from \texttt{POR} to \texttt{Normal} mode, when the physician sends the command to change to \texttt{Normal} mode, i.e., the event \texttt{@T( mCommand=NORMAL)} occurs, but the battery voltage still is low. This error can be fixed adding an extra condition when a command is received from the DCM. We then refine some rows of the mode transition table as follows:

{\scriptsize
\texttt{
\begin{center}
\begin{tabular}{|l|l|l|}
\hline
OLD MODE & EVENT & NEW MODE\\
\hline
POR & @T( mCommand=NORMAL ) when & Normal\\
    & mBATTERYvoltage $\geq$ BatteryLevel & \\
\hline
POR & @T( mCommand=TMP ) when 	& Temporal\\
    & mBATTERYvoltage $\geq$ BatteryLevel	 & \\
\hline
\end{tabular}
\end{center}
}
}

\noindent
Regarding scenario \textit{S2}, the violation shows a case where the Pacemaker is in \texttt{MAGnormal} mode and the monitored variable \texttt{mMODEbrad} changes to \texttt{off}. Then, when the magnet is released, the system could not enter to \texttt{Normal} mode because no program is set. We read again the informal description and we noticed that the description of scenario \textit{S2} was too general. If the scenario considers engaging the \texttt{Normal} mode when the magnet is released and a program is set in \texttt{mMODEbrad}, the tool does not find violations.

We now assess the efficiency of Spin to analyze the previous scenarios. We measure the time for generating counterexamples for each scenario and verifying the properties (once the specification is fixed).  We compare the results with the approach presented in~\cite{DBLP:conf/fase/AguirreFMMW09}, which uses the DynAlloy Analyzer for analyzing these scenarios. The results are summarized in Table~\ref{tab:comparison-spin-dynalloy} (we refer to our tool as \textit{Spin+}).  All experiments were run in an Intel Core 2 Duo of 2.26Ghz processor with 4GB of RAM, running Mac OS X.

\begin{table}[h]
{\small
\begin{center}
\begin{tabular}{ c | c | c | c | c |}
\cline{2-5}
& \multicolumn{2}{|c|}{Counterexamples} & \multicolumn{2}{|c|}{Verification}\\
\cline{2-5}
\cline{2-5}
& time & depth/lurs & time & depth/lurs \\
\hline
\multicolumn{1}{|c|}{Spin+ (\textit{S1})} & 90ms &  10.000 depth & 2.450ms /  18.900ms & 10.000 /  1.000.000 depth\\
\hline
\multicolumn{1}{|c|}{DynAlloy (\textit{S1})} & 9771ms & 5  & 42.222ms /  $>$30min & 10 / 20 lurs\\
\hline
\multicolumn{1}{|c|}{Spin+ (\textit{S2})} & 780ms & 10.000 depth  & 1.160ms /  7.700ms &  10.000 /  1.000.000 depth\\
\hline
\multicolumn{1}{|c|}{DynAlloy (\textit{S2})} & 899ms & 3 &  12.301ms / 18.154ms / 167.130ms & 20 /  30 / 50 lurs\\
\hline
\end{tabular}
\end{center}
}
\caption{Performance comparison between our tool and DynAlloy.}%
\label{tab:comparison-spin-dynalloy}
\end{table}

Notice that both Spin and DynAlloy Analyzer are able to find the counterexamples. However, for verification tasks, the performance of DynAlloy Analyzer gets worse as the loops unrolls (lurs) are increased. It needs more than 30' for verifying the scenario \textit{S1} considering 20 lurs.

\section{Conclusion}
\label{conclusion}

We have presented a tool that provides a way of specifying and analyzing particular sets of (critical) executions of SCR specifications. The tool involves an operational notation that allows the engineer to describe behavioural scenarios by means of programs. These programs are composed of sentences that represent arbitrary or restricted changes in the state, and tests, that can be combined using sequential composition and iterations. Furthermore, the tool is able to automatically translate the scenarios to Promela so that they can be analyzed via Spin, an efficient \emph{off-the-shelf} model checker of free availability.

We applied the SCR method to a Pacemaker system and we used its formal tabular requirements specification as a running example. The selection of this case study was motivated by the Formal Methods Challenge of the Software Quality Research Laboratory of McMaster University~\cite{Pacemaker-Spec}. We carried out some experiments, to assess the ability of the tool in finding problems in the specification, and its efficiency compared with a previous approach, based on DynAlloy. 

The notation for scenarios is limited to safety properties. While this was enough for DynAlloy, which is only able to analyze bounded executions, our use of model checking as a backend analysis tool enables the possibility of also verifying liveness properties. We are then working on extending the notation for scenarios, to include the possiblity of adequately specifying liveness properties.

\bibliographystyle{eptcs}
\bibliography{generic}

\end{document}